# A Simple Observer for Gyro and Accelerometer Biases in Land Navigation Systems

Vasiliy M. Tereshkov

(*Topcon Positioning Systems LLC, Moscow, Russia*)

(Email: VTereshkov@topcon.com)

In various applications of land vehicle navigation and automatic guidance systems, Global Navigation Satellite System/Inertial Measurement Unit (GNSS/IMU) positioning performance crucially depends on the attitude determination accuracy affected by gyro and accelerometer bias instabilities. Traditional bias estimation approaches based on the Kalman filter suffer from implementation complexity and require non-intuitive tuning procedures. In this paper we propose, as an alternative, a simple observer that estimates inertial sensor biases exclusively in terms of quantities with obvious geometrical meaning. By this, any multidimensional vector-matrix operations are avoided and observer tuning is substantially simplified. The observer has been successfully tested in a farming vehicle navigation system.

KEY WORDS

1. GNSS. 2. IMU. 3. Kalman Filter. 4. Bias Estimation

1. INTRODUCTION. The advent of compact and inexpensive Micro-Electromechanical Sensors (MEMS) has led to the development of consumer-grade Inertial Measurement Units (IMU). These devices have proved to be applicable in many areas unreachable for classical inertial instruments. One of the successfully addressed problems is the navigation and automatic control of various land vehicles. Whereas the primary positioning device for such vehicles is usually a receiver of a Global Navigation Satellite System (GNSS), the desired control performance is achieved only if this receiver is integrated with an IMU or other sensors.

This problem is highlighted in some specific applications such as precision agriculture where the aim is to guide an automated farming machine along a swath with minimum cross-track displacement. The control problem is usually formulated for the "reference point" located at the centre of gravity or in the middle of the rear or the front axle, depending on the vehicle kinematics and the implement used. On the other hand, the GNSS receiver is installed atop the machine cabin roof. Therefore, the position and velocity provided by the receiver are substantially different from those observed at the "reference point". The differences depend on the continuously varying attitude angles that cannot be measured by a single GNSS receiver (Thuilot et al., 2002; Jongmin, 2014). There appears a need for an IMU-based Attitude and Heading Reference System (AHRS) integrated with the GNSS receiver (Topcon Precision Agriculture, 2014; John Deere, 2014).

Among the error sources of such an integrated system, gyro and accelerometer biases play a key role. MEMS devices are known to have large and unpredictable run-to-run and in-run bias instabilities. For slope compensation in precision agriculture problems, only the longitudinal and lateral gyros and accelerometers are of interest, as they directly determine the accuracy of attitude computations (Salychev, 2004).

Consider a simple numerical example. Let the vertical displacement of the GNSS receiver from the vehicle "reference point" be $h = 3\,\text{m}$, and the vehicle roll angle caused by the terrain slope be $\varphi = 5\,°$. If this slope is neglected, it results in a cross-track positioning error of $h \sin \varphi = 26\,\text{cm}$. It is a dozen times larger than the GNSS inherent errors in the most precise Real-Time Kinematic (RTK) navigation mode. Now suppose we combine this GNSS receiver with a MEMS AHRS to compensate for the slope, but the lateral accelerometer has a run-to-run bias of $b_a = 0.2\,\text{m/s}^2$. This bias causes a roll computation error of $\tilde{\varphi} = b_a / g = 1.2\,°$ (where we assume $g = 9.8\,\text{m/s}^2$), so the corresponding positioning error will be $h \sin \tilde{\varphi} = 6\,\text{cm}$. It still exceeds typical RTK errors, so the use of RTK is meaningless until the accelerometer bias is estimated.

A traditional way to estimate gyro and accelerometer biases on-line is to apply the Kalman filter (Farrell and Barth, 1999; Salychev, 2004; Ding et al., 2007). The AHRS is treated as a multidimensional dynamical system with external measurements provided by its GNSS receiver and IMU. Some of its state variables are identified with the attitude angles, some others with the unknown gyro and accelerometer biases. Using the Kalman framework, all the state variables are estimated as long as their observability conditions are satisfied.

The observability issue is very non-trivial. While gyro bias estimation does not require any specific vehicle manoeuvres, accelerometer biases are completely unobservable in a straight motion as they cannot be separated from the attitude errors. To make the estimation possible, the vehicle should perform a series of turns (Salychev, 2004). This is an inherent property of any AHRS and does not depend on the chosen estimation technique. Fortunately, a typical path of a land vehicle is composed of straight segments and turns, so the estimation of accelerometer biases can be successfully performed.

Besides the issues associated with the physical essence of the AHRS operation, there is a practical problem with the Kalman filter itself. The universality of this estimation method is achieved at the cost of implementation complexity. The filter requires an efficient framework for manipulating multidimensional, often sparse, matrices and vectors. This is especially challenging for embedded computer systems with very limited hardware resources. Moreover, Kalman filter performance crucially depends on the *a priori* information. The designer should specify three covariance matrices: one for the initial state uncertainty, and two others for the process noise and the measurement noise. At least one of them, the process noise covariance matrix, has no clear physical meaning and cannot be deduced from the sensor characteristics available to the designer (Salychev, 2004). Therefore, in most engineering applications some laborious and non-intuitive iterative procedures are needed to tune the Kalman filter. For the same reason, the Kalman filter optimality, though guaranteed by theory, is rarely achieved in practice.

2. GEOMETRIC OBSERVERS. The search for a less demanding estimation method than the Kalman filtering has led to the invention of a group of techniques which we call geometric observers. What they all have in common is that the estimates of all desired variables are expressed only in terms of quantities with clear geometrical meaning. These observers make use of three-dimensional vectors in the conventional Euclidean space, and no multidimensional vector-matrix operations are needed.

2.1. *Attitude Estimation*. Geometric observers for GNSS/IMU integrated navigation have been developed from two completely different standpoints. The first approach (Shaw et al., 1981; Salychev, 2012) is based on the deep similarities of error behaviour in gimballed and strapdown inertial devices. To emphasise this fact, a notion of a "mathematical", or "virtual", gimballed platform is introduced to study the operation of a strapdown AHRS. The direction cosine matrix, updated by the AHRS computer, therefore describes the orientation of the vehicle body with respect to this "virtual platform". Furthermore, any attitude errors are treated as angular deviations of the "virtual platform" plane from the true local level plane. To correct the errors, a control angular rate vector is applied to the "virtual platform" unless these two planes coincide. This control angular rate replaces the correction term of the Kalman filter.

The second standpoint is the group theory, from which a family of so-called "invariant" observers can be derived (Bonnabel and Rouchon, 2005; Mahony et al., 2008). This research area is very attractive and fruitful, yet the practical results are the same as obtained by the "virtual platform" technique.

2.2. *Bias Estimation*. It is natural to equip a geometric attitude observer with a gyro bias estimator (Mahony et al., 2008). In the steady-state operation, the averaged control angular rate applied to compensate the effects of the gyro bias becomes equal and opposite to that bias. This immediately determines the necessary observer structure. Nevertheless, to the best of our knowledge, no one has succeeded in incorporating an accelerometer bias estimator into the same observer. The only known attempt (Grip et al., 2011) seems to be impractical since it relies on the squared norm of the measured specific force vector to determine its bias. This leads to doubtful performance on noisy data.

In the following sections we will derive the unified estimator for gyro and accelerometer biases in the context of a geometric attitude observer. The outline of this proposal (Tereshkov, 2013) suffered from some unnecessary restrictions imposed on the vehicle trajectory. In the more general theory presented here, these restrictions are removed and a more feasible estimation technique is developed.

3. BIAS OBSERVER DESIGN. We start the observer construction with several assumptions that are reasonable in the context of precision agriculture applications:

A) The IMU is subjected to factory calibration, so the residual sensor biases are small, as well as the attitude errors caused by these biases.

B) Farming machines typically move on nearly flat terrain, so the roll and pitch angles and the corresponding angular rates are small; vertical accelerations are negligible compared with gravity.

C) The vehicle does not suffer from sideslip and the GNSS signals are not blocked by any obstacles, so the heading angle is perfectly known as the direction of the vehicle velocity measured by the GNSS receiver.

D) The effects of Earth's rotation and curvature are not taken into consideration.

Now, if we neglect all the products of at least two small values, we can arrive at a linearized error dynamics model, from which the bias observer structure can be deduced and its stability proven. Any possible violations of our assumptions, e.g., during the motion on a hill slope, will not necessarily impair the practical performance of the observer but will of course invalidate its theoretical justifications.

A serious challenge in the observer construction is to decide what available quantities are suitable to drive the observer. The most obvious choice is to use the raw measurements provided by the IMU and the GNSS receiver. Unfortunately, they are coupled with the sensor biases through several complicated differential equations (Titterton and Weston, 2004), and this choice returns us to the implementation problems we wished to avoid. Then, following the guidelines of the geometric observers theory, we select the control angular rate to drive the bias observer. So we now need to find how this angular rate is coupled with the quantities of interest, i.e., with gyro and accelerometer biases.

3.1. *Attitude Dynamics.* Consider a strapdown AHRS that computes the direction cosine matrix $\mathbf{C}$ between the body frame $b = \{x_b, y_b, z_b\}$ and the "virtual platform" frame $p = \{x_p, y_p, z_p\}$. The body frame is chosen such that $x_b$ points forwards, $y_b$ rightwards, $z_b$ downwards. The "virtual platform" frame is slightly misaligned from the North-East-Down frame $n = \{N, E, D\}$.

The matrix dynamics are described by the equation (Salychev, 2004)

$$\dot{\mathbf{C}} = \mathbf{C}\breve{\boldsymbol{\omega}}_b^b - \breve{\boldsymbol{\omega}}_p^p \mathbf{C} \qquad (1)$$

Here a subscript denotes the frame for which the angular rate is specified, and the superscript denotes the frame to which the rate is projected. Thus, $\boldsymbol{\omega}_b^b$ is the body frame angular rate as measured by the gyros, $\boldsymbol{\omega}_p^p$ is the "virtual platform" control angular rate as seen from this "virtual platform". An arc denotes a skew-symmetric matrix corresponding to a vector, so that $\breve{\mathbf{a}}\mathbf{b} = \mathbf{a} \times \mathbf{b}$.

Let us project the accelerometer measurements $\mathbf{f}_b$ onto the "virtual platform":

$$\mathbf{f}_p = \mathbf{C}\mathbf{f}_b \qquad (2)$$

For a perfect AHRS installed on a vehicle that moves on a flat non-rotating Earth (Assumption D), the "virtual platform" lies in the local level plane, so that $\mathbf{f}_p = \mathbf{f}_n$. Therefore, the first two components of $\mathbf{f}_p = [f_N, f_E, f_D]^T$ contain only the actual accelerations, but not the projections of the gravity vector $\mathbf{g}$. If this ideal condition is violated and the "virtual platform" is not horizontal, there appears a difference between the measured specific force $\mathbf{f}_p$ and the true specific force $\mathbf{f}_n^{GNSS} = \dot{\mathbf{v}}_n^{GNSS} - \mathbf{g}$. The latter is provided by differentiating the vehicle velocity $\mathbf{v}_n^{GNSS}$ measured by the GNSS receiver. The difference is fed back to the "virtual platform" in the form of the control angular rate:

$$\boldsymbol{\omega}_p^p = -\breve{\mathbf{k}}_p (\mathbf{f}_n^{GNSS} - \mathbf{f}_p) \qquad (3)$$

Here the vertical vector $\mathbf{k} = -k\mathbf{g}/g$ represents the attitude correction gain, and the quantity $\tau = 1/(kg)$ is the correction time constant. Since in the perfect case $\boldsymbol{\omega}_p^p = \mathbf{0}$, any variation of this angular rate is equal to the rate itself: $\delta\boldsymbol{\omega}_p^p = \boldsymbol{\omega}_p^p$.

3.2. *Attitude Error Dynamics.* In the presence of gyro and accelerometer biases, $\mathbf{b}_g$ and $\mathbf{b}_a$, the "virtual platform" tilt error $\boldsymbol{\theta}$ appears. This error is small due to Assumption A, so we are able to represent it by a vector and express the direction cosine matrix variation as $\delta \mathbf{C} = \breve{\boldsymbol{\theta}}_p^p \mathbf{C}$.

By varying Equation (1) and substituting $\delta \boldsymbol{\omega}_b^b = \mathbf{b}_g$ and $\delta \boldsymbol{\omega}_p^p = \boldsymbol{\omega}_p^p$, we can get

$$\dot{\boldsymbol{\theta}}_p^p = \mathbf{C}\mathbf{b}_g - \boldsymbol{\omega}_p^p \qquad (4)$$

Equation (4) has a clear physical meaning. The rate of change of the attitude error vector is determined by two opposite factors: the gyro biases (projected onto the "virtual platform") and the control angular rate. However, this equation is not convenient for the analysis of strapdown AHRS performance, since its terms are not constant during the vehicle turns. Instead, they are modulated by the $\mathbf{C}$ matrix, even if the sensor biases, $\mathbf{b}_g$ and $\mathbf{b}_a$, are constant by themselves. To avoid this problem, Equation (4) can be transformed to the body frame:

$$\dot{\boldsymbol{\theta}}_b^b + \breve{\boldsymbol{\omega}}_b^b \boldsymbol{\theta}_b = \mathbf{b}_g - \boldsymbol{\omega}_p^b \qquad (5)$$

It is the last term $\boldsymbol{\omega}_p^b = \mathbf{C}^T \boldsymbol{\omega}_p^p$ that will be used for the estimation of inertial sensor biases. The relation between $\boldsymbol{\omega}_p^b$ and $\mathbf{b}_g$ is already given by Equation (5). To establish a similar relation between $\boldsymbol{\omega}_p^b$ and $\mathbf{b}_a$, we vary Equation (2) with $\delta \mathbf{f}_b = \mathbf{b}_a$:

$$\delta \mathbf{f}_p = \breve{\boldsymbol{\theta}}_p^p \mathbf{f}_p + \mathbf{C}\mathbf{b}_a \qquad (6)$$

Then we substitute Equation (6) into Equation (3):

$$\boldsymbol{\omega}_p^p = \breve{\mathbf{k}}_p \breve{\boldsymbol{\theta}}_p^p \mathbf{f}_p + \breve{\mathbf{k}}_p \mathbf{C}\mathbf{b}_a = \left(\mathbf{k}_p^T \mathbf{f}_p\right) \boldsymbol{\theta}_p - \left(\mathbf{k}_p^T \boldsymbol{\theta}_p\right) \mathbf{f}_p + \breve{\mathbf{k}}_p \mathbf{C}\mathbf{b}_a \qquad (7)$$

The first term in the right-hand side of Equation (7) is proportional to the attitude error $\boldsymbol{\theta}$ we wish to mitigate. It provides the desired error feedback that justifies our choice of the control law (Equation (3)). According to Assumption B, vertical accelerations are small, so $f_D = -g$ and $\mathbf{k}_p^T \mathbf{f}_p = kg$. The second term is an additional error that appears when the vectors $\mathbf{k}$ and $\boldsymbol{\theta}$ are not orthogonal. Due to Assumption C, there is no heading error and only attitude errors are present. Then, $\boldsymbol{\theta}$ lies in the level plane, while $\mathbf{k}$ is always vertical by its definition, so the whole term vanishes. The third term reflects the influence of accelerometer biases on the attitude correction accuracy. Though generally harmful, this term allows for the estimation of accelerometer biases by means of the $\boldsymbol{\omega}_p^b$ signal. As stated in Assumption B, roll and pitch angles are small, so we can freely replace $\mathbf{k}_b$ by $\mathbf{k}_p$ and treat this vector as constant in any vehicle manoeuvres.

After accepting these simplifications and transforming Equation (7) to the body frame, we finally get

$$\boldsymbol{\omega}_p^b = kg\, \boldsymbol{\theta}_b + \breve{\mathbf{k}}_p \mathbf{b}_a \qquad (8)$$

Equations (5) and (8) completely determine the attitude error dynamics and relate it to the sensor biases and to the "virtual platform" control angular rate.

3.3. *Control Rate Dynamics*. The attitude error by itself is neither measurable nor important for estimation. From our control law (Equation (3)) we know that, once sensor biases are estimated and compensated, the attitude error tends to zero. So it is expedient to eliminate this quantity from the final equations. First, we differentiate Equation (8), provided that $\mathbf{b}_a$ is constant, and get $\dot{\boldsymbol{\omega}}_p^b = kg\,\dot{\boldsymbol{\theta}}_b^b$. Second, we substitute Equation (8) and its derivative into Equation (5):

$$\dot{\boldsymbol{\omega}}_p^b = -\left(\breve{\boldsymbol{\omega}}_b^b + kg\mathbf{I}\right)\boldsymbol{\omega}_p^b + kg\mathbf{b}_g + \breve{\boldsymbol{\omega}}_b^b \breve{\mathbf{k}}_p \mathbf{b}_a \tag{9}$$

The vector triple product in the last term of Equation (9) can be significantly simplified if we utilise the Assumption B and set $\boldsymbol{\omega}_b^b = [0, 0, \omega_{zb}]^T$. Then, $\breve{\boldsymbol{\omega}}_b^b \breve{\mathbf{k}}_p \mathbf{b}_a = k\omega_{zb}\mathbf{b}_a$, so the product is collinear with $\mathbf{b}_a$. Further, we recall the definition of the attitude correction time constant $\tau = 1/(kg)$ and rewrite Equation (9) in the form

$$\tau\dot{\boldsymbol{\omega}}_p^b = -\left(\tau\breve{\boldsymbol{\omega}}_b^b + \mathbf{I}\right)\boldsymbol{\omega}_p^b + \mathbf{b}_g + \omega_{zb}\mathbf{b}_a / g \tag{10}$$

The eigenvalues of the dynamics matrix $-\left(\tau\breve{\boldsymbol{\omega}}_b^b + \mathbf{I}\right)$ are $-1$ and $-1 \pm i\tau\omega_{zb}$. Therefore, the solution of Equation (10) is always stable.

This is exactly what we desired to obtain. Equation (10) describes a linearized coupling between sensor biases and control angular rate applied to the "virtual platform". We will use it to justify the bias observer structure.

3.4. *Bias Estimate Dynamics*. The individual observers for gyro and accelerometer biases will be both driven by the control angular rate $\boldsymbol{\omega}_p^b$ multiplied by some scalar gains. The choice of these gains should respect the observability conditions imposed by the nature of the system under study. In the Introduction we noticed that accelerometer biases are completely unobservable in a straight motion. This conclusion is confirmed by Equation (10). As far as $\omega_{zb} = 0$, the control angular rate $\boldsymbol{\omega}_p^b$ is by no means affected by accelerometer biases. Therefore, we will set the accelerometer bias observer gain proportional to the vehicle angular rate $\omega_{zb}$.

Our observer equations finally take the form:

$$\dot{\hat{\mathbf{b}}}_g = \boldsymbol{\omega}_p^b / \tau_g \tag{11}$$

$$\dot{\hat{\mathbf{b}}}_a = g\omega_{zb}\boldsymbol{\omega}_p^b / \omega_a \tag{12}$$

Here, $\tau_g$ can be called the gyro bias observer time constant, and $\omega_a$, by analogy, the accelerometer bias observer "rate constant". These quantities play a role of adjustable parameters of the observers. They determine the settling time and the noise level of the obtained estimates.

At this stage, having derived the general observer equations, we can construct the estimation block diagram (Figure 1).

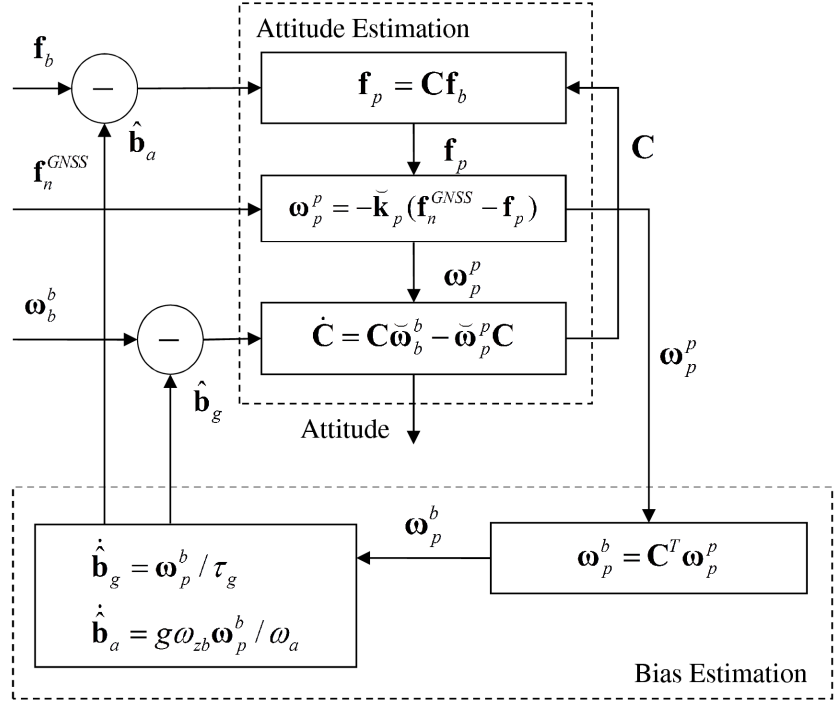

Figure 1. Bias Observer Block Diagram.

As seen from Equation (3), the bias observer requires the derivative of the measured velocity $\mathbf{v}_n^{GNSS}$. It is generally admitted that numerical differentiation, when applied to real-world signals, results in large noise levels. Nevertheless, GNSS receivers are capable of providing quite smooth velocity measurements based on carrier phase raw data, even if the receiver position is updated using much noisier pseudorange information (Farrell and Barth, 1999). Since the observer does not rely on position measurements, its accuracy is enough for our particular purposes (Section 4).

*3.5. Stability.* When the current bias estimates are fed back to the input of the attitude observer, as shown in Figure 1, the residual sensor biases affecting the attitude accuracy are expressed as $\mathbf{b}_g = \mathbf{b}_g(0) - \hat{\mathbf{b}}_g$ and $\mathbf{b}_a = \mathbf{b}_a(0) - \hat{\mathbf{b}}_a$. The complete linearized closed-loop dynamics of these biases is determined by Equations (10) – (12) and admits the trivial zero solution. To prove its stability, we can construct a positive-definite candidate Lyapunov function:

$$V = \frac{\tau}{2}\left|\boldsymbol{\omega}_p^b\right|^2 + \frac{\tau_g}{2}\left|\mathbf{b}_g\right|^2 + \frac{\omega_a}{2g^2}\left|\mathbf{b}_a\right|^2 \qquad (13)$$

Its time derivative according to dynamics equations is

$$\dot{V} = -\boldsymbol{\omega}_p^{bT}\left(\tau\breve{\boldsymbol{\omega}}_b^b + \mathbf{I}\right)\boldsymbol{\omega}_p^b \qquad (14)$$

This quadratic form is negative-definite due to the specific properties of the matrix $-\left(\tau\breve{\boldsymbol{\omega}}_b^b + \mathbf{I}\right)$ pointed out in Section 3.3. Therefore, $V$ approaches its minimum value at the origin as long as $\mathbf{f}_p \neq \mathbf{f}_n^{GNSS}$, and the trivial solution of Equations (10) – (12) is stable.

3.6. *Steady-State Operation*. Suppose that the vehicle is moving straight. The open-loop solution $\boldsymbol{\omega}_p^b$ of Equation (10) will eventually converge to its steady value such that $\dot{\boldsymbol{\omega}}_p^b = \mathbf{0}$. Equation (10) will then take the simplest possible form

$$\mathbf{b}_g = \boldsymbol{\omega}_p^b \qquad (15)$$

This result is in agreement with what was said in Section 2.2. The control angular rate merely compensates the gyro bias and can be directly used to estimate it.

Now suppose that the gyro bias has already been estimated and compensated, and the vehicle is turning at some *constant* angular rate $\omega_{zb}$. Then from the steady-state solution of Equation (10) we get:

$$\mathbf{b}_a = \frac{g}{\omega_{zb}}\left(\tau\breve{\boldsymbol{\omega}}_b^b + \mathbf{I}\right)\boldsymbol{\omega}_p^b = g\begin{bmatrix} 1/\omega_{zb} & -\tau & 0 \\ \tau & 1/\omega_{zb} & 0 \\ 0 & 0 & 0 \end{bmatrix}\boldsymbol{\omega}_p^b \qquad (16)$$

Unfortunately, Equation (16), first obtained in our previous paper (Tereshkov, 2013) without considering the general observer Equations (10) – (12), is of very limited engineering importance. Indeed, the steady-state conditions are satisfied only when the vehicle angular rate $\omega_{zb}$ is maintained constant for a time interval three to five times longer than the attitude correction time constant $\tau$. Typical turns performed by land vehicles are significantly shorter, so the application of Equation (16) is not justified by practice. The general observer seems to be preferable over its elegant but unreliable steady-state solutions.

4. EXPERIMENTAL VALIDATION. The bias observer discussed in Section 3.4 was implemented in the special experimental firmware of a Topcon AGI-4 GNSS/IMU integrated receiver. The receiver was installed on the cabin roof of a MacDon 9300 self-propelled windrower (Figure 2). The windrower was driven along a curved path (Figure 3) so that the observability of both gyro and accelerometer biases was achieved.

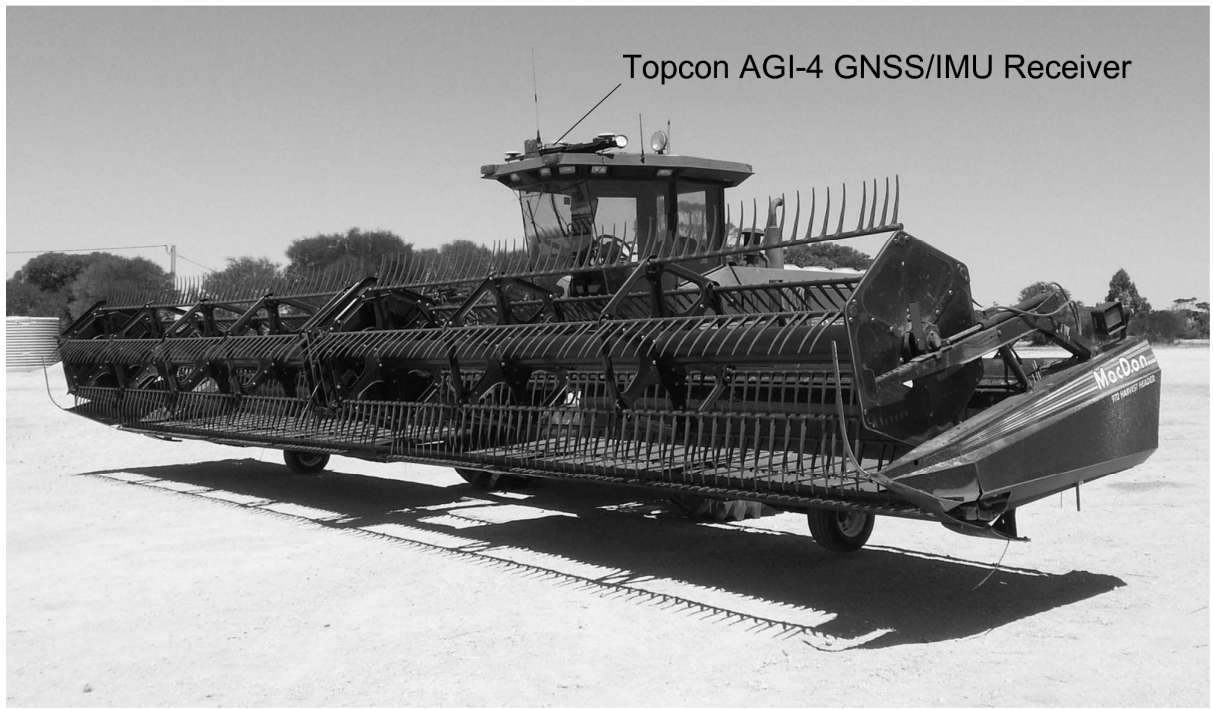

Figure 2. MacDon 9300 Self-Propelled Windrower with Topcon AGI-4 GNSS/IMU Receiver.

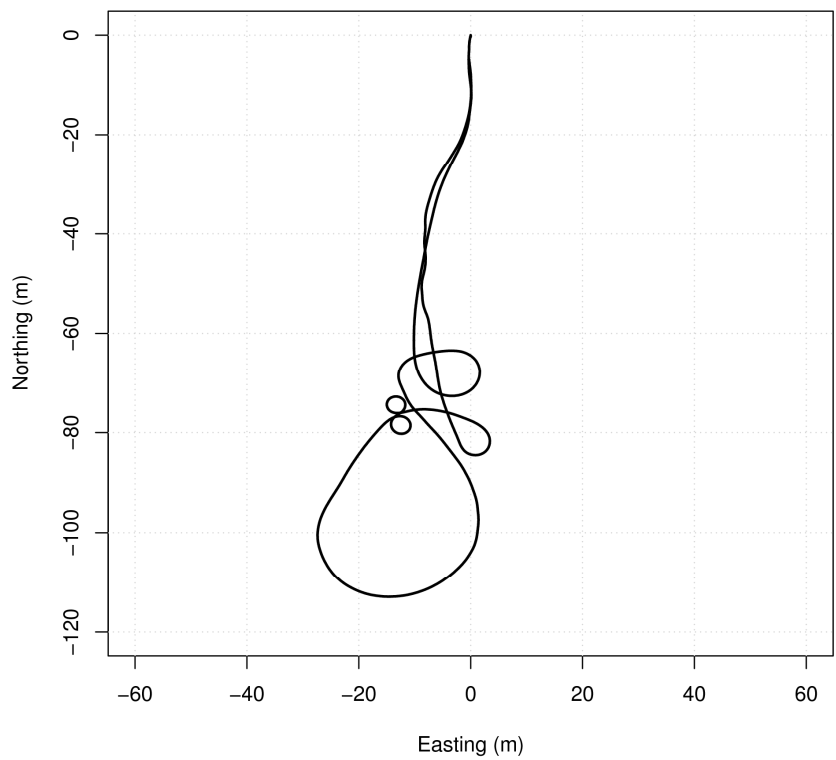

Figure 3. Test Path.

To assess the observer accuracy, test run data were post-processed in two stages. At the first stage, the actual gyro and accelerometer biases for the longitudinal and lateral axes were estimated by the proposed observer and then compensated. Bias values were found to be $b_{gx} = -0.08 \text{ deg/s}$, $b_{gy} = -0.13 \text{ deg/s}$, $b_{ax} = -0.11 \text{ m/s}^2$, $b_{ay} = 0.00 \text{ m/s}^2$, which is typical for the consumer-grade MEMS sensors installed in the AGI-4 receiver. Nevertheless, these values by themselves cannot give any information regarding the bias observer performance. Since any other in-run bias estimation methods could have their own performance issues, none of them could serve as a reliable reference. For that reason, the second data processing stage was needed: the precisely known artificial biases were added to the sensor measurements, and the estimation procedure was repeated. The resulting bias estimates are shown in Figures 4 and 5.

Data post-processing was performed with the following observer parameter values: $\tau = 4 \text{ s}$, $\tau_g = 40 \text{ s}$, $\omega_a = 45 \text{ deg/s}$. The values were selected manually, but, in contrast to Kalman filter tuning, this was a straightforward task, as the bias observer performance is completely determined by these three constants with a rather clear and natural meaning (Section 3.4).

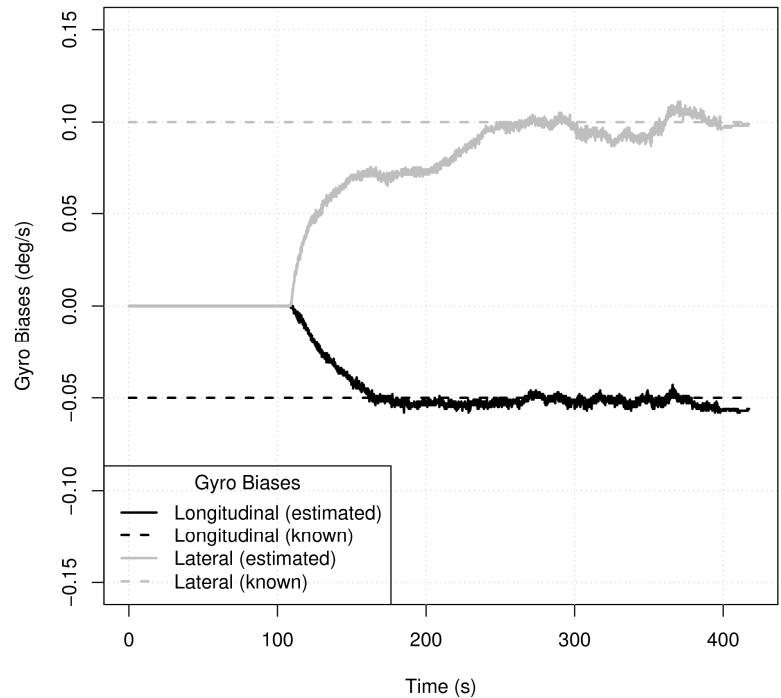

Figure 4. True and Estimated Gyro Biases.

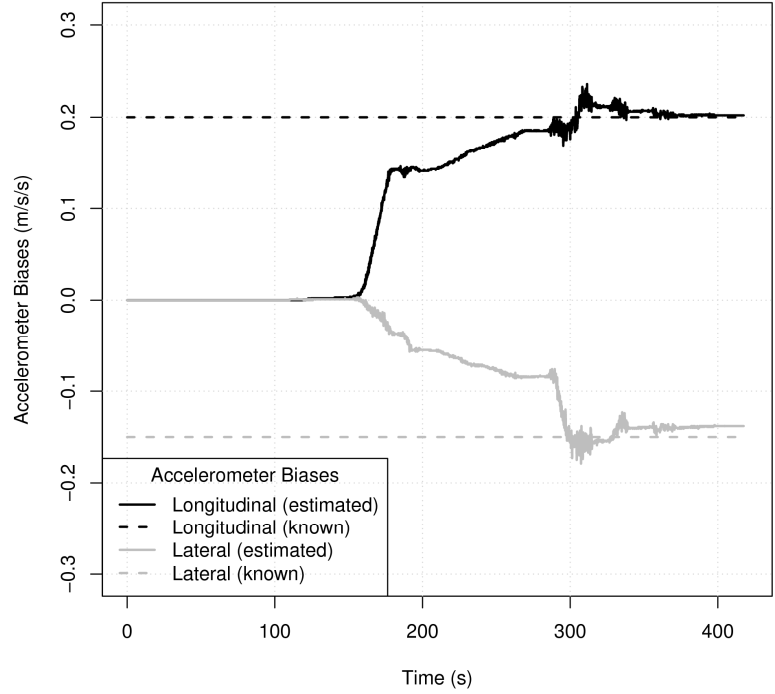

Figure 5. True and Estimated Accelerometer Biases.

After the initial convergence is complete, the residual estimate variations are of the order of $\tilde{b}_g = 0.01\,\text{deg/s}$ and $\tilde{b}_a = 0.01\,\text{m/s}^2$, which can be attributed partly to observer imperfections and partly to the actual uncompensated in-run bias instabilities correctly tracked by the observer. Consider the worst case when the estimate variations are solely due to observer errors. Let the vertical displacement of the GNSS/IMU integrated receiver from the vehicle "reference point" be $h = 3\,\text{m}$, as in the example discussed in the Introduction. Then, the "reference point" positioning errors caused by the wrong attitude determination are $h\tilde{b}_g \tau = 2\,\text{mm}$ and $h\tilde{b}_a / g = 3\,\text{mm}$ respectively. It is by one order of magnitude less than the GNSS receiver positioning errors even in the RTK mode.

The bias observer proved to be quite robust against parameter variations: a 20 % change in any of the three adjustable constants leads to bias estimate deviations smaller than $0.001\,\text{deg/s}$ and $0.01\,\text{m/s}^2$.

5. CONCLUSIONS. Motivated by the excessive implementation complexity of the traditional Kalman filter and inspired by the geometric observer theory, we sought for a simple observer that could estimate gyro and accelerometer biases of a MEMS-based GNSS/IMU integrated land navigation system. Our primary aim was to express the desired bias estimates exclusively in terms of scalars and vectors in the conventional three-dimensional Euclidean space, thus avoiding any multidimensional vector-matrix operations.

The designed bias observer (Figure 1) was successfully tested on a self-propelled windrower (Figures 4 and 5). The achieved estimation accuracy proved to be sufficient for all precision agriculture tasks.

It is still unclear whether the proposed observer can be generalised to applications other than land vehicle navigation, since the typical dynamics of aerial or underwater vehicles violates the Assumptions B and C made when deriving the observer equations. A complete geometrical treatment of the bias estimation problem is a subject of future work.

FINANCIAL SUPPORT. This research received no specific grant from any funding agency, commercial or not-for-profit sectors.

CONFLICT OF INTEREST. None.